\documentclass[twocolumn,floatfix,prl,aps,showpacs]{revtex4-1}
\usepackage{graphicx,amsmath,amssymb,color}

\usepackage[titletoc,title]{appendix}
\usepackage{hyperref}
\usepackage{gensymb}
\usepackage{lineno}
\usepackage{nicefrac}

\newcommand{\be}{\begin{equation}}
\newcommand{\ee}{\end{equation}}

\newcommand{\ba}{\begin{eqnarray}}
\newcommand{\ea}{\end{eqnarray}}

\def\bea{\begin{eqnarray}}
\def\eea{\end{eqnarray}}
\def\ba{\begin{equation}\begin{array}{c}}
\def\ea{\end{array}\end{equation}}
\def\be{\ba\displaystyle}
\def\ee{\ea}

\begin{document}

\title{Even denominator fractional quantum Hall states in higher Landau levels of graphene}
\author{Youngwook Kim$^{1,\star}$, Ajit C.~Balram$^{2,\star}$, Takashi Taniguchi$^{3}$, Kenji Watanabe$^{3}$, Jainendra K.~Jain$^{4}$, Jurgen H. Smet$^{1}$}
\affiliation{$^{1}$Solid State Nanophysics, Max Planck Institute for Solid State Research, Heisenberstra\ss e 1, D-70569 Stuttgart, Germany}
\affiliation{$^{2}$Niels Bohr International Academy and the Center for Quantum Devices, Niels Bohr Institute, University of Copenhagen, 2100 Copenhagen, Denmark}
\affiliation{$^{3}$National Institute for Material Science, 1-1 Namiki, Tsukuba, 305-0044, Japan}
\affiliation{$^{4}$Department of Physics, 104 Davey Lab, Pennsylvania State University, University Park, Pennsylvania 16802, USA}
\affiliation{$^{*}$These authors have contributed equally.}
\date{\today}
\maketitle
{\bf An important development in the field of the fractional quantum Hall effect has been the proposal that the 5/2 state observed in the Landau level with orbital index $n = 1$ of two dimensional electrons in a GaAs quantum well~\cite{Willett87} originates from a chiral $p$-wave paired state of composite fermions which are topological bound states of electrons and quantized vortices. This state is theoretically described by a ``Pfaffian" wave function~\cite{Moore91} or its hole partner called the anti-Pfaffian~\cite{Levin07,Lee07}, whose excitations are neither fermions nor bosons but Majorana quasiparticles obeying non-Abelian braid statistics~\cite{Read00}. This has inspired ideas on fault-tolerant topological quantum computation~\cite{Nayak08} and has also instigated a search for other states with exotic quasiparticles. Here we report experiments on monolayer graphene that show clear evidence for unexpected even-denominator fractional quantum Hall physics in the $n=3$ Landau level. We numerically investigate the known candidate states for the even-denominator fractional quantum Hall effect, including the Pfaffian, the particle-hole symmetric Pfaffian, and the 221-parton states, and conclude that, among these, the 221-parton appears a potentially suitable candidate to describe the experimentally observed state. Like the Pfaffian, this state is believed to harbour quasi-particles with non-Abelian braid statistics~\cite{Wen91}.}

The fractional quantum Hall effect (FQHE) has generated some of the most exotic emergent states in condensed matter. While the finest FQHE is seen in high quality GaAs quantum wells, graphene offers the possibility to discover new ground states since the electron-electron interactions in $n\neq 0$ Landau levels (LLs) of graphene are different from those in GaAs [see section III.1 of Supplementary Information (SI)]. A large number of fractional quantum Hall states have been observed in the $n=0$ LL~\cite{Feldman12,Feldman13}, primarily at fractional fillings $\nu=s/(2ps\pm 1)$, where $s$ and $p$ are positive integers. These fractional quantum Hall states are understood as $\nu^*=s$ integer quantum Hall states of composite fermions carrying $2p$ vortices~\cite{Jain89}. Experiments on high quality bilayer graphene have revealed several even denominator states~\cite{Ki14,Kim15,Zibrov16,Li17}, which are believed to be analogous to the 5/2 state of GaAs systems and the even denominator states observed in ZnO~\cite{Falson15,Falson18}. More recently, even denominator fractions have been reported in the $n=0$ LL of graphene~\cite{Zibrov18}, whose origin is not yet fully understood. Progress has also been made toward identifying the topological content of the 5/2 state in GaAs through thermal Hall measurements, which show evidence for Majorana edge states but are not consistent with either the Pfaffian or the anti-Pfaffian model~\cite{Banerjee17}. Here we address the discovery of fractional quantum Hall physics at half filling of the $n=3$ LL of monolayer graphene, where the standard composite fermion physics is no longer considered viable.

\begin{figure*}[!tbh]
\begin{center}
\includegraphics{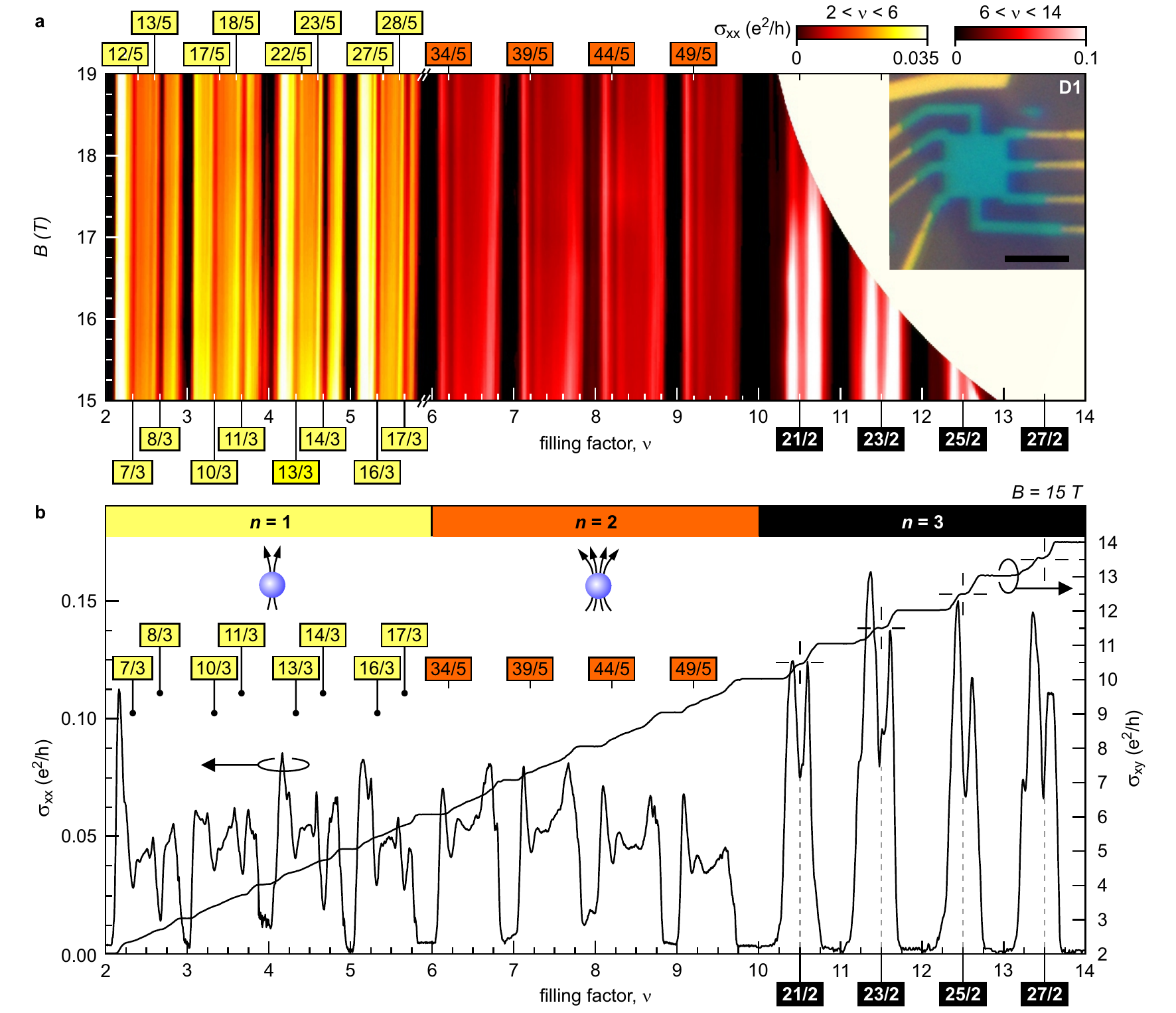}
\caption{{\bf Longitudinal ($\sigma_{\rm xx}$) and Hall conductivity ($\sigma_{\rm xy}$) of sample D1.} {\bf a,} Color rendition of the longitudinal conductivity as the $n = 1$, 2 and 3 Landau levels get gradually filled by tuning the electron density with the graphite backgate. Colored boxes mark filling factors at which the electronic system apparently condenses in an incompressible fractional quantum Hall state. To distinguish the details in the data across the full dynamic range of the conductivity, two color scales have been introduced valid below and above filling factor 6 respectively. The inset shows an optical image of the device. {\bf b,} Line traces of the longitudinal and Hall conductivity recorded at 15 T. The orbital index of the Landau level being filled is marked at the top. All displayed data were recorded at approximately 30 mK.}
\label{Fig1}
\end{center}
\end{figure*}

\begin{figure}[!tbh]
\begin{center}
\includegraphics{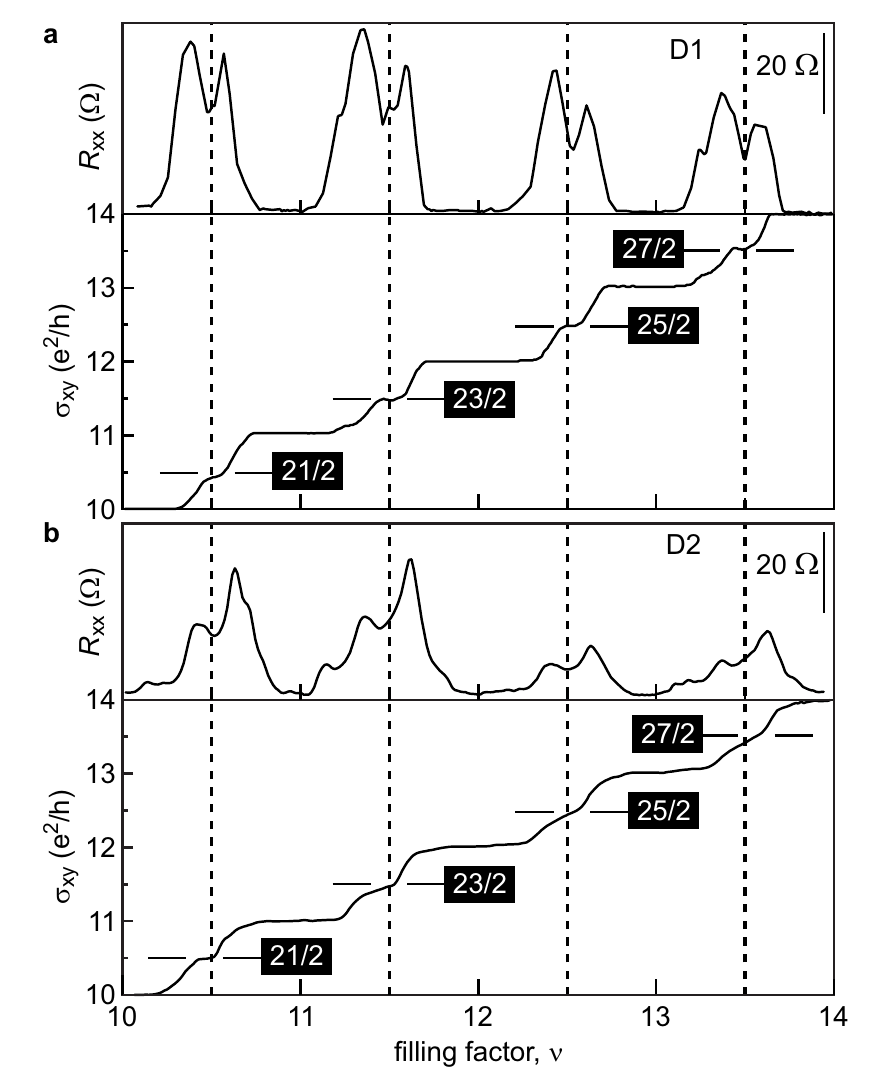}
\caption{{\bf Hall plateaus at half filling of the $n=3$ levels.} {\bf a}, Data recorded on  device D1 for $B = 15\ {\rm T}$. {\bf b}, Same for device D2 and $B = 21.5\ {\rm  T}$. All data have been acquired at a temperature of approximately 30 mK.}
\label{Fig2}
\end{center}
\end{figure}

The experimental investigations have been carried out on van der Waals heterostructures composed of a graphene monolayer sandwiched between thicker hexagonal boron nitride (hBN) layers. Graphite back or front gates offered density tunability. Details of the sample fabrication are deferred to the Methods section. Fig.~\ref{Fig1} gives an overview of the Hall and longitudinal conductivity recorded at 30 mK and a magnetic field of 15 T. By gradually raising the electron density, the Landau levels with orbital indices 1, 2 and 3 get successively filled. Both the valley and spin degeneracy of these levels have been fully lifted as apparent from the observed integer quantum Hall states.

In the $n = 1$ LL incipient fractional quantum Hall states are observed at 1/3, 2/5, 2/3 and 3/5.  These states follow the conventional sequence of composite fermions with two quantized vortices attached, and have been anticipated in samples of sufficiently high quality. They were also observed in state-of-the-art samples reported in the literature~\cite{Dean11,Amet15}. Their manifestation serves as a clear quality indicator. More evidence for outstanding sample quality follows from  signatures of the FQHE at filling 1/5 in the $n = 2$ LL in Fig.~\ref{Fig1}. It is attributed to four flux composite fermion sequence of fractional quantum Hall states. Even stronger examples for holes are seen in Fig.~S3 of SI.

Finally, new fractional quantum Hall territory is charted when the $n = 3$ LL gets partially occupied. Clear minima arise in $\sigma_{\rm xx}$ whenever one of the four spin and valley split levels with orbital index 3 is half filled. In panel {\bf b} of Fig.~\ref{Fig1}, these minima are accompanied by well developed plateaus in the Hall conductivity $\sigma_{xy}$ with properly quantized values equal to the conductance quantum $e^{2}/h$ multiplied by $\bar{\nu} + 1/2$. Here, $e$ is the electron charge, $h$ the Planck constant and $\bar{\nu}$ the number of completely filled valley and spin resolved LLs. An enlarged view of the data has been replotted in Fig.~\ref{Fig2} together with an additional data set acquired on a second van der Waals heterostructure D2 in panel {\bf b}. This second device also exhibits clear features of even denominator fractional quantum Hall physics, although a fully developed plateau is only available for the lowest valley and spin split $n = 3$ level. These incompressible states are fragile. They disappear quickly with increasing temperature and require the largest magnetic fields. Typical dependencies on both parameters are contained in section III of SI together with an attempt to extract the thermal activation energy (section II of SI). This fragility contrasts with the two-flux and four-flux composite fermion states in the $n = 1$ and $n = 2$ LLs which persist even up to 1.7 K and fields down to 6 T (section I of SI).

The appearance of even-denominator fractional quantum Hall states in the $n = 3$ LL comes entirely unanticipated. In conventional semiconductor based two-dimensional electron systems spontaneous symmetry breaking and charge density wave physics in the form of stripe and bubble phases enters the scene in the $n = 1$ LL, where it still competes with the FQHE~\cite{Lilly99b,Pan99,Xia04}. However, in higher LLs the charge density wave physics subdues the fractional quantum Hall physics altogether. Apparently, the balance of power is shifted in graphene  because of the  spinor nature of the LL wave functions which decisively alters the effective electron-electron interaction in $n\neq 0$ LLs (section IV.1 of SI). For instance, recent theoretical work has highlighted that as a result charge density wave phases are not expected in the $n = 1$ LL in monolayer graphene but their first appearance should move to higher LLs compared to GaAs~\cite{Knoester16}.  The experimental data at hand forces theory to revisit the viability of fractional quantum Hall behavior in the $n = 3$ LL and to investigate the nature of the observed incompressible states at half filling of the spin and valley split $n=3$ LLs.

Before addressing the even denominator states in more detail, it is instructive to consider the standard odd-denominator states. Consistent with previous theory and experiments~\cite{Dean11,Feldman12,Feldman13,Amet15,Balram15c}, both two- and four-flux composite fermions are stabilized in the $n=0$ and $n=1$ LLs. Which fractional quantum Hall state is stabilized depends on the interaction matrix elements.  For electrons confined to a LL, the interaction is fully specified by the Haldane pseudopotentials $V_m$~\cite{Haldane83}, which are the energies of a pair of electrons in states with relative angular momentum $m$ (section IV.1 of SI).  These pseudopotentials enable the transformation of the problem to solve to a mathematically equivalent one of electrons residing in the lowest LL and interacting through an effective interaction that has the same Haldane pseudopotentials as the Coulomb interaction in the partially filled $n$th LL. The actual interelectron interaction is modified in a complex manner due to screening by gates and also because of LL mixing. It is expected that screening will make the interaction more short ranged, i.e.~it will effectively increase the $V_1$ pseudopotential relative to others, while LL mixing will decrease all $V_m$. The corrections due to LL mixing, parametrized by the dimensionless quantity $\kappa=(e^2/(\epsilon\ell)/(\hbar v_{\rm F} \ell))$, where $v_{\rm F}$ is the Fermi velocity, have been considered in a perturbative approach~\cite{Peterson13}.  The value of $\kappa$ varies from 2.2 for suspended graphene to 0.5-0.8 for graphene on hBN~\cite{Peterson13}. We find that while the bare $n=2$ LL Coulomb interaction produces $\nu=1/5$ FQHE, consistent with our experimental observation, the modified interaction given in Ref.~\cite{Peterson13} with $\kappa=0.5-0.8$ fails to do so, suggesting that, while valid at small $\kappa$, the modified interaction overestimates the deviation from pure Coulomb at the experimental values of $\kappa$. We therefore consider a range of interactions in the vicinity of the $n$th LL Coulomb interaction by allowing some variation of the pseudopotentials $V_1$ and $V_3$, which we will denote as $\delta V_{\rm 1}$ and $\delta V_{\rm 3}$. We note that only pseudopotentials $V_m$ with odd $m$ are relevant for single component electrons.
 As discussed in section IV.2 of SI, theoretical comparisons with the exact Coulomb eigenstates indicate that only four-flux composite fermions are stabilized in the $n=2$ LL, and neither two- nor four-flux composite fermions are anticipated in the $n=3$ LL. This is consistent with the fact that in the $n=2$ LL the 1/5 FQHE is observed while the 1/3 FQHE is absent, and in the $n=3$ LL there is no sign of either the 1/3 or the 1/5 state.

Numerical calculations, summarized in section III.3 of SI, reveal that in the $n = 3$ LL, neither the Pfaffian nor the particle-hole symmetric Pfaffian have a good overlap with the exact ground state in the vicinity of the pure Coulomb interaction point. Various two-component states were also tested and found to not be viable in the $n=3$ LL (section IV.4 of SI). We therefore appeal to the parton paradigm~\cite{Jain89b}, a generalization of the composite fermion construction, which can also serve as a source of inspiration to construct potentially relevant Ansatz wave functions for incompressible ground states. In the parton framework each electron is first decomposed into a set of fictitious particles referred to as partons. Each parton species is then placed into some integer quantum Hall state and finally the partons are fused back together to recover the physical electron. The parton construction obtains the standard states of composite fermions, but also more general states that do not lend themselves to an obvious interpretation in terms of composite fermions, some of which also support non-Abelian excitations~\cite{Wen91}. Here, we consider the so-called 221-parton state with the wave function~\cite{Jain89b,Wen91,Wu16}
\be
\Psi^{\rm 221-parton}={\cal P}_{\rm LLL}\Phi_2^2\Phi_1,
\ee
where $\Phi_n$ is the wave function of $n$ filled LLs. ${\cal P}_{\rm LLL}$ denotes the projection into the  lowest LL, which we evaluate by calculating the overlap of the unprojected wave function with each lowest LL basis function with the Monte Carlo technique~\cite{Balram16b}; this approach yields explicit lowest LL wave functions for up to $N=12$ particles. A definitive realization of the 221-parton state has not yet been achieved, although it has been theoretically proposed for a certain parameter regime in bilayer graphene~\cite{Wu16}.

\begin{figure}[tb!]
\begin{center}
\includegraphics{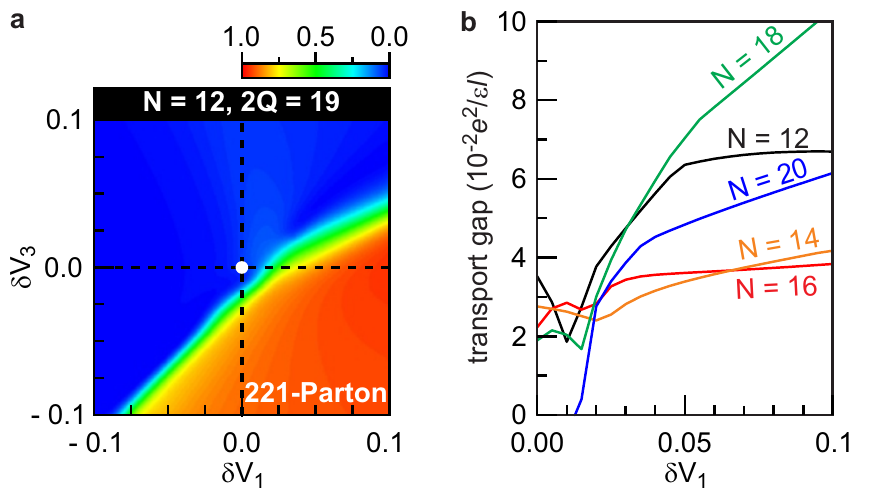}
\caption{{\bf Overlaps and transport gaps at half filling of the $n=3$ Landau level.} {\bf a}, Overlap for the 221-parton state. Simulations were performed for a $41 \times 41$ grid of parameter pairs. The white dot in the center of the  panel marks the exact Coulomb point in the $n=3$ graphene LL and $\delta V_1$ and $\delta V_3$ denote changes to the first two relevant Haldane pseudopotentials.
{\bf b},  Transport gap for $n=3$ Landau level of graphene extracted from exact diagonalization in the vicinity of the Coulomb interaction. The transport gap is shown for 21 values of the interaction defined by $V_m+\delta V_m$, where $V_m$ are the Haldane pseudopotentials for the pure Coulomb interaction, $\delta V_1=-\delta V_3$, and all other $\delta V_m=0$. }
\label{Fig3}
\end{center}
\end{figure}

Encouragingly, the 221-parton state exhibits a large overlap across a substantial area of parameter space with the pure Coulomb interaction point close to the perimeter of this region (see Fig.~\ref{Fig3}{\bf a}). In order to assess further whether it is realistic for this parton state to appear in experiment, it is necessary to demonstrate a non-vanishing transport gap, which is the energy required to create a pair of far separated quasi-particles with charge $\pm e/4$. These transport gaps are displayed in Fig.~\ref{Fig3}{\bf b} for systems of 12 to 20 particles. They are obtained from exact diagonalization of the model Hamiltonian along a diagonal line extending from the Coulomb point into the red region in Fig.~\ref{Fig3}{\bf a}. It has unfortunately not been possible to obtain reliable estimates for the gap in the thermodynamic limit, because of strong finite size fluctuations, a common feature of delicate fractional quantum Hall states in higher Landau levels~\cite{Morf02}.  Nonetheless, the transport gap remains stable in a large part of the interaction space and is on the order of 0.02 $e^2/(\epsilon \ell)$ near the Coulomb point. Hence, deviations from the Coulomb point of only a few \% in the interaction pseudopotentials are sufficient to stabilize the 221-parton state in the $n=3$ graphene LL, making it possible that the experimentally observed state is adiabatically connected and thus topologically equivalent to the 221-parton wave function. A convincing proof, however, will require a more accurate account of the  interaction between electrons including all corrections. This is an inherently complex problem beyond the scope of this study.

The 221-parton state is topologically distinct from the Pfaffian, anti-Pfaffian and the particle-hole symmetric Pfaffian states, which, at least in principle, enables experiments to distinguish between them. While all these states have the same quasi-particle charge and Ising non-Abelian statistics, they differ in other properties, such as the quasi-particle tunneling exponent and the number of backward moving neutral modes, as reviewed in Ref.~\cite{Banerjee17}. In particular, the 221-parton state has a thermal Hall conductance of $\kappa_{xy}=5/2$ in units of $(\pi^2 k_{\rm B}^{2}/(3h))T$, whereas the Pfaffian, particle-hole symmetric Pfaffian and anti-Pfaffian states have, in the same units, $\kappa_{xy}=3/2$, $1/2$ and $-1/2$, respectively  (in addition to the contribution from the filled LLs).  The three states also have different Hall viscosities, given by~\cite{Read09} $\eta_{\rm H} = \hbar \mathcal{S}/(16\pi\ell^2)$, where $\mathcal{S}$ is the ``shift"~\cite{Wen92} in the spherical geometry~\cite{Haldane83} equal to $3$, $-1$, $1$ and $5$ for the Pfaffian, anti-Pfaffian, particle-hole symmetric Pfaffian and the 221-parton states, respectively. Since we are working with only two-body interactions, our study respects particle-hole symmetry in the LL in question and thus treats the 221 parton state and its hole partner as equivalent, but LL mixing can break the tie to select one of them. The hole partner of the 221 parton state has $\kappa_{xy}=-3/2$ and $\mathcal{S}=-3$. The 221-parton may also be viewed as a manifestation of  $f$-wave pairing of composite fermions~\cite{Balram18}.

 While we have presented compelling experimental evidence for the existence of an unanticipated incompressible  fractional quantum Hall state in state-of-the-art encapsulated monolayer graphene when Landau levels with orbital index 3 are half filled, a decisive validation of the 221-parton state requires further theoretical and experimental work. It is conceivable that the 221-parton state is the first state in the parton sequence (22n) and the next member of this sequence (222) which occurs at 2/3 filling or its particle hole conjugate at 1/3 filling would possibly be seen as the sample quality improves further. \\

\bibliography{biblio_fqhe_Nature}
\bibliographystyle{naturemag}

\ \linebreak
\noindent {\bf Acknowledgments}
We acknowledge useful discussions with K. von Klitzing, Inti Sodemann, Ying-Hai Wu, and Jun Zhu, and assistance for sample preparation by S. G\"{o}res and M. Hagel. We extend our gratitude to S. Masubuchi and T. Machida for fruitful input on the ELVACITE stamp method for the fabrication of van der Waals heterostructure.  J.H.S.~is grateful for financial support from the graphene flagship. Y.K. thanks the Humboldt Foundation and A.C.B. the Villum Foundation for support. The Center for Quantum Devices is funded by the Danish National Research Foundation. This project has received funding from the European Research Council (ERC) under the European Union's Horizon 2020 research and innovation programme [grant agreement number: 1104931001 (TOPDYN)]. The work at Penn State was supported by the U. S. Department of Energy under Grant no. DE-SC0005042. Some portions of this research were conducted with Advanced CyberInfrastructure computational resources provided by The Institute for CyberScience at The Pennsylvania State University. Some of the numerical calculations were performed using the DiagHam package, for which we are grateful to its authors. The growth of hexagonal boron nitride crystals was supported by the Elemental Strategy Initiative conducted by the MEXT, Japan and the CREST (JPMJCR15F3), JST.\linebreak
\ \linebreak
\noindent {\bf Author Contributions}
The experiments have been designed by Y.K and J.H.S. They were carried out in the laboratory by Y.K.
The theory was performed by A.C.B and J.K.J. The calculations were run by A.C.B.
T.T. and K.W. synthesized the h-BN bulk crystal.
Y.K., A.C.B., J.K.J, and J.H.S. contributed to the manuscript writing. \linebreak
\ \linebreak
\noindent{\bf Author Information}
 Reprints and permissions information is available at
www.nature.com/reprints. The authors declare no competing financial
interests. Readers are welcome to comment on the online version of the paper.
Correspondence and
requests for materials should be addressed to J.H.S. (j.smet@fkf.mpg.de)

\noindent {\large {\bf METHODS}}

\noindent {\bf Device fabrication.} The investigations in this work have been carried out on four different devices referred to as D1 through D4. D1 is a van der Waals heterostructure composed of a graphene monolayer sandwiched between thicker hBN layers and this stack is placed on top of a graphite back gate. Devices D2 to D4 are in addition equipped with a top graphite gate covered by a hBN multilayer. Devices D1 and D2 were manufactured under ambient condition, whereas devices D3 and D4 were stacked in vacuum  ($5\times10^{-4}$ mbar). Except for the vacuum environment, all procedures for stacking are the same. Here we briefly describe the fabrication steps for these four devices.

D1 has been assembled with the help of a modified viscoelastic stamping method~\cite{Gomez14}. The top hBN layer was directly exfoliated to a commercially available viscoelastic stamp (Gel-Pak, PF-30/17-X4). A suitable hBN flake was carefully selected in the dark field image of an optical microscope according to the following criteria: thickness homogeneity, absence of wrinkles and bubbles and the presence of a corner with an angle of $120\degree$ indicative of flake termination along the main crystal directions. The selected flake had a thickness of approximately 10 nm. The subsequent graphene and hBN layer of the van der Waals heterostructure were first exfoliated on top of a 90 nm thick SiO$_{2}$ thermally grown on a Si substrate. A graphene and hBN layer with straight boundaries and a $120\degree$ corner were chosen. The bottom BN had at thickness of about 20 nm.

The viscoelastic method offers control over the pick-up and release process simply by varying the contact area between the SiO$_2$ substrate and the layers already available on the stamp. For instance, if the top hBN layer directly exfoliated on the stamp touches the substrate completely, hBN will be transferred to the substrate. However, if the hBN covers entirely a graphene flake on the substrate and only partially touches the substrate, the graphene will be picked up instead. To increase the yield during pick-up, the hBN initially exfoliated on the stamp is chosen at least 2 times bigger than the graphene and bottom hBN layers that need to be picked up. A typical size of this hBN layer is about $40~{\rm \mu m} \times  40~{\rm \mu m}$.

Using an appropriate tool offering $x,y,z$ as well as rotational motion, the monolayer graphene to be picked up was slowly approached by the stamp holding the top hBN layer. To avoid an emerging Moir\'{e} superlattice potential to our device, the boundaries of the two flakes were aligned by aiming for a $5\degree$ to $30\degree$ twist angle between the hBN and the graphene flake. Upon establishing contact, the graphene was finally picked up. Following the same procedure, the bottom hBN and bottom graphite were picked-up. The heterostructure was then transferred onto a Si substrate with a 280 nm thick SiO$_2$ surface layer. During the entire pick-up procedure, the sample stage was kept at a temperature of 120\degree C. In order to expand the useable area of the sample with atomic scale flatness, the stack was annealed at 500\degree C in forming gas at an ambient pressure of 150 mbar for 30 min. We did not observe any changes in the relative orientation and placement of the constituent flakes as a result of the annealing~\cite{Wang15c}.  Electron beam lithography was deployed for etching and deposition of contacts. The details of these processing steps can be found elsewhere~\cite{Kim16}.

The fabrication of devices D2-D4 is more complex as it requires two additional pick-up steps in order to add a front gate and encapsulation with hBN.  To accomplish this successfully and with reasonable yield, we resorted to the same basic procedures but with an alternative, polymer-based  stamp rather than a viscoelastic stamp~\cite{Wang13,Masubuchi18}. Devices D2-D4 were annealed under identical conditions as Device D1. For device D4, a  Moir\'{e} superlattice potential was imposed by choosing a zero degree alignment between the graphene monolayer and the adjacent hBN layer covering it from the top. To avoid multiple  Moir\'{e} superlattice potentials, the bottom hBN was intentionally misaligned by  $30\degree$ with respect to the already picked up hBN/graphene heterostructure.

The hBN layer was etched in a CHF$_3$/O$_2$ or SF$_6$/Ar plasma. The slow etching speed for carbon enables the selective removal of the hBN layer on top of the graphite. the latter is etched with the help of an O$_2$ plasma which leaves hBN intact. Hence, the middle hBN safely protects graphene during the etching of the graphitic top gate. Finally, the middle hBN layer, the graphene layer as well as the bottom hBN layer are removed either with a CHF$_3$/O$_2$ or SF$_6$/Ar plasma. To avoid a short between the graphene and the bottom graphite gate, the bottom hBN layer is only partially etched away.\linebreak
\noindent {\bf Transport measurements.}
The transport measurements on D1-D3 were recorded in a top-loading-into-mixture dilution refrigerator (Oxford Instruments) at a base temperature of approximately 30 mK. D4 was investigated in a physical property measurement system of Quantum Design (PPMS Dynacool) down to temperatures of 1.7 K.  Transport measurements were performed with conventional four terminal lock-in techniques at an excitation current $I$ which varied between 10 and 100 nA at a frequency of 17.777 Hz.\linebreak
\noindent{\bf Data availability.} The data that support the findings of this study are available from the corresponding author on reasonable request.

\end{document}